\documentclass[aps, reprint, superscriptaddress, longbibliography]{revtex4-2}

\usepackage{graphicx}
\graphicspath{{images/}}

\usepackage{bm}
\usepackage{dsfont}
\usepackage[dvipsnames]{xcolor}
\usepackage{pstricks}
\usepackage{verbatim}
\usepackage{units}
\usepackage{natbib}
\usepackage{mathtools}
\usepackage{multirow}
\usepackage{enumitem}
\usepackage{mathrsfs}
\usepackage{leftidx}
\usepackage{xspace}
\usepackage{amsmath}
\usepackage[normalem]{ulem}
\usepackage{soul}
\usepackage{physics}
\usepackage{tabularx}
\usepackage{upgreek}
\usepackage{times}

\usepackage[normalem]{ulem}
\newcolumntype{Y}{>{\centering\arraybackslash}X}

\usepackage{hyperref}
\hypersetup{colorlinks=true,linktoc=all,linkcolor=blue,breaklinks=true,citecolor=blue,urlcolor=blue}

\usepackage{color}

\usepackage[english]{babel}
\usepackage{subfiles}

\addto\captionsenglish{}

\newcommand{\subfigref}[2]{\ref{#1}\hyperref[#1]{(#2)}}
\newcommand{\Vs}{V_\text{SD}}
\newcommand{\VKILL}{V_\text{KILL}}
\newcommand{\VPROBE}{V_\text{PROBE}}
\newcommand{\VPUMP}{V_\text{PUMP}}
\renewcommand{\L}{\text{L}}
\newcommand{\R}{\text{R}}
\newcommand{\LR}{\text{L,R}}

\newcommand{\In}{\text{in}}
\newcommand{\Out}{\text{out}}
\renewcommand{\L}{\text{L}}
\newcommand{\Gb}{\Gamma_\text{tot}}
\newcommand{\G}{\Gamma}
\newcommand{\GL}{\Gamma_\text{L}}
\newcommand{\GR}{\Gamma_\text{R}}

\renewcommand{\arraystretch}{1.3}

\makeatletter
\renewcommand\paragraph{\@startsection{paragraph}{4}{\z@}%
  {3.25ex \@plus1ex \@minus.2ex}%
  {-0em}%
  {\normalfont\normalsize\itshape\indent}}

\begin{abstract}

Self-oscillations are the result of an efficient mechanism generating periodic motion from a constant power source. In quantum devices, these oscillations may arise due to the interaction between single electron dynamics and mechanical motion. We show that, due to the complexity of this mechanism, these self-oscillations may irrupt, vanish, or exhibit a bistable behaviour causing hysteresis cycles. We observe these hysteresis cycles and characterize the stability of different regimes in both single and double quantum dot configurations. In particular cases, we find these oscillations stable for over 20 seconds, many orders of magnitude above electronic and mechanical characteristic timescales, revealing the robustness of the mechanism at play.

\end{abstract}

\begin{document}

\title{Stability of long-sustained oscillations induced by electron tunneling\\
}

\author{Jorge Tabanera-Bravo}
\affiliation{Dept.~Estructura de la Materia, F\'isica T\'ermica y Electr\'onica and GISC, Universidad Complutense de Madrid. 28040 Madrid, Spain}

\author{Florian Vigneau}
\affiliation{Department of Materials, University of Oxford, Oxford OX1 3PH, United Kingdom}

\author{Juliette Monsel}
\affiliation{Department of Microtechnology and Nanoscience (MC2), Chalmers University of Technology, S-412 96 G\"oteborg, Sweden\looseness=-1}

\author{Kushagra Aggarwal}
\affiliation{Department of Materials, University of Oxford, Oxford OX1 3PH, United Kingdom}

\author{L\'ea Bresque}
\affiliation{Univ.~Grenoble Alpes, CNRS, Grenoble INP, Institut N\'eel, 38000 Grenoble, France}

\author{Federico Fedele}
\affiliation{Department of Materials, University of Oxford, Oxford OX1 3PH, United Kingdom}

\author{Federico Cerisola}
\affiliation{Department of Materials, University of Oxford, Oxford OX1 3PH, United Kingdom}
\affiliation{Physics and Astronomy, University of Exeter, Exeter EX4 4QL, United Kingdom}

\author{G.A.D. Briggs}
\affiliation{Department of Materials, University of Oxford, Oxford OX1 3PH, United Kingdom}

\author{Janet Anders}
\affiliation{Physics and Astronomy, University of Exeter, Exeter EX4 4QL, United Kingdom}
\affiliation{Institut f\"ur Physik und Astronomie, Potsdam University, 14476 Potsdam, Germany}

\author{Alexia Auff\`eves}
\affiliation{MajuLab, CNRS-UCA-SU-NUS-NTU International Joint Research Laboratory}
\affiliation{Centre for Quantum Technologies, National University of Singapore, 117543 Singapore, Singapore}

\author{Juan M.R. Parrondo}
\affiliation{Dept.~Estructura de la Materia, F\'isica T\'ermica y Electr\'onica and GISC, Universidad Complutense de Madrid. 28040 Madrid, Spain}

\author{Natalia Ares}
\email{natalia.ares@eng.ox.ac.uk}
\affiliation{Department of Engineering Science, University of Oxford, Oxford OX1 3PJ, United Kingdom}

\date{\today}

\maketitle

\setlength{\tabcolsep}{4pt}
\renewcommand{\arraystretch}{1.3}

\section{Introduction}

Coupling of quantum devices to mechanical degrees of freedom can be exploited for high-precision measurements \cite{feng2020precision,pirkkalainen2015cavity,rodrigues2019coupling, Moser2013UltrasensitiveFD, de2018ultrasensitive, o2010quantum} and may serve as a platform for quantum and classical information processing \cite{lake2021processing, aporvari2021strong}. At low temperatures, carbon nanotube (CNT) devices can be operated as extremely sensitive mechanical oscillators which are strongly coupled to single electron tunneling~\cite{huttel2009carbon, huttel2010single, lassagne2009coupling,Meerwaldt2012,sazonova2004tunable,vigneau2021, wang2021, tavernarakis, Moser2013UltrasensitiveFD, Chaste2012ANM,bachtold2022Review}. The interplay between single electron tunneling and mechanical motion, in the absence of a mechanical drive, can give rise to self-sustained oscillations. Such self-oscillations were observed to be either present or absent depending on the electron transport regime, both in theory ~\cite{blanter2004single, usmani_strong_2007,Bennett2006Nov} and experiments ~\cite{steele2009,Bachtold2011SelfOsc,Huttel2012,Huttel2015,Laird2020,Bachtold2020,Baugh2020}.
{\black In this paper, we report that, at the boundary between different
 electron transport regimes, self-oscillations can appear or vanish spontaneously. These bistable states of motion of an undriven oscillator, until now unexplored, are of particular interest for applications of these devices in quantum and stochastic thermodynamics \cite{strasberg2, wachtler2019stochastic,elouard2015reversible, monsel2018autonomous}, in collective dynamics and synchronization \cite{wachtler2020}, and in emulating neural behavior \cite{lin_vanadium_2018, stoliar_implementation_2021, rocco_exponential_2022}.
We} measure bistable self-oscillations both for single and double quantum dot configurations and present a theoretical analysis that provides a complete characterization of the stability of self-oscillations at different bias voltages. We find that, once started, undriven oscillations can be self-sustained and may decay over time scales of the order of $10^8$ mechanical periods.  In this way, our system explores timescales of electronic and mechanical origin that are separated by several orders of magnitude.

\section{Experiments and model}

\begin{figure}[h!]
    \centering
    \includegraphics[scale = 1]{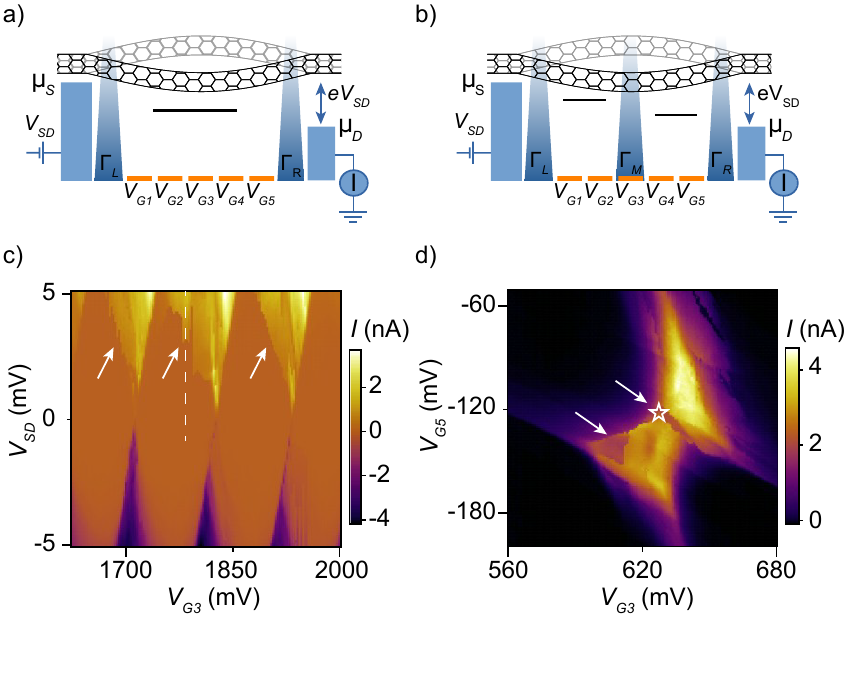}

    \caption{a,b) Schematic of the device in single dot (a) and double dot (b) configurations. A CNT is suspended between the source and drain electrodes. Five gate voltages, $V_\text{G1-G5}$, are used to create either a single dot or a double dot. A bias voltage $\Vs$ drives a current $I$ through the CNT. The chemical potentials of the source and drain electrodes are  $\mu_\text{S}=e\Vs$ and $\mu_\text{D}=0$, respectively. The right, left and interdot tunnel rates are indicated and the associated tunnel rates are labeled $\Gamma_\R$, $\Gamma_\L$ and $\Gamma_{\rm M}$.
    c) Current measured in single-dot configuration as a function of the gate voltage $V_\text{G3}$ and bias voltage $\Vs$. White arrows point at features which indicate the presence of self-oscillations. The current traces in Fig.~\subfigref{fig:SD}{a} were taken along the dashed vertical line.
    d) Current measured in the double-dot configuration by sweeping  $V_\text{G5}$ and stepping $V_\text{G3}$ with $\Vs = 1.8$~mV. White arrows point at current features which indicate the presence of self-oscillations. The white star indicates a triple point, where we observe the switch in current as a function of $\Vs$ plotted in Fig.~\subfigref{fig:DD}{a}.
    }
    \label{figdevice}
\end{figure}
Our electromechanical device consists of a fully suspended CNT (see Fig.~\subfigref{figdevice}{a}) ~\cite{Laird2016,Laird2018,Laird2020,vigneau2021}. Applying a bias voltage $\Vs$  between the source and drain electrodes, we measure a current $I$ through the CNT.
The gate electrodes, to which we apply voltages $V_\text{G1-5}$, are located beneath the CNT. These gate voltages define an electrostatic potential for the confined charges, and depending on the combination of gate voltage values, we can define single or double quantum dots within the CNT (Figs.~\subfigref{figdevice}{a,b})~\cite{laird_quantum_2015, hanson2007spins}.

The mechanical motion of the CNT can be excited by applying a radio-frequency (rf) signal to one of the gate electrodes. Sharp changes in the current through the CNT indicate its mechanical motion~\cite{huttel2009carbon,Bachtold2011,laird2012}, and we identify the natural mechanical resonance frequency $\Omega/2\pi=270$~MHz {\color{black} with an approximate quality factor $Q$ of 2000 (see Supplemental Material~\cite{suppl}). Experiments were performed at 60 mK}. At charge degeneracy points, strong coupling between electron tunnelling and mechanical motion is evidenced by the softening of the mechanical resonance frequency~\cite{lassagne2009coupling, steele2009,huttel2010single,Meerwaldt2012,Ilani2014,vigneau2021}.

Self-oscillations are identified as sharp switches in current appearing in the absence of an rf excitation~\cite{steele2009,Bachtold2011SelfOsc,Huttel2012,Huttel2015,Laird2020,Baugh2020}.  We observe such switches both for single and double dot configurations (Fig.~\subfigref{figdevice}{c,d}).
While in a double-dot configuration, these sharp switches are visible within the bias triangles (Fig.~\subfigref{figdevice}{d}).

\subsection{Single-dot configuration}
\begin{figure}[h!]
  \centering
  \includegraphics[scale = 1]{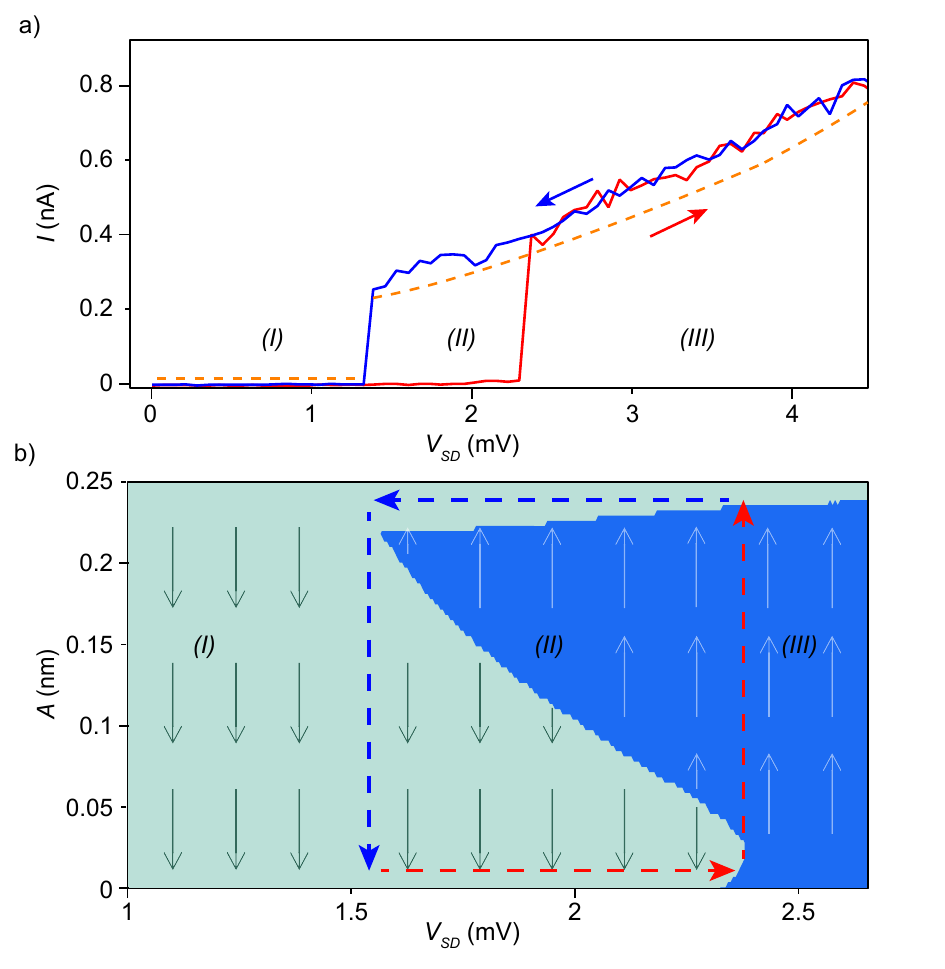}
  \caption{
  a) Current switch hysteresis as a function of $\Vs$ measured in the single dot regime following the dashed line ($V_\text{G3}=1.8$~V) in  Fig.~\subfigref{figdevice}{c}. The red and blue arrows indicate the direction of each current sweep. The orange dashed line is the current calculated from Eq.~\eqref{eq:current}.  The numerals (I, II, III) indicate regions of no oscillations, bistability and self-oscillations, respectively.
  b) Stability diagram: the dark (light) blue areas indicate $\Delta E > 0$ $(\Delta E <0)$ for different values of $A$ and $\Vs$. Small arrows  indicate the direction in which $A$ would change in each area given the sign of $\Delta E$. Blue and red arrows delimit regions (I), (II) and (III) and indicate a similar hysteresis cycle as that shown in panel a). We use $\Omega/2\pi=270$ MHz, $g_{\rm m}=0.01$~eV$/$nm, $m =2\times 10^{-22}$ kg, and $Q = 1000$, $\Gamma_\L=400~\text{ GHz}$, $\Gamma_\R=18~\text{ GHz}$,  $\alpha_\L=4~\text{ nm}^{-1}$, and  $\alpha_\R=0$.
  \label{fig:SD}
  }
\end{figure}

In the single-dot configuration, we observe the bistability through the hysteretic behavior of the current when sweeping $\Vs$ in and out of the self-oscillation area, following the vertical white dashed line in Fig.~\subfigref{figdevice}{c}, which  corresponds to a constant gate voltage $V_\text{G3}=1.8$~V.
As $\Vs$ is swept upward, a sharp increase of current around $\Vs=2.3$~mV indicates the onset of self-oscillations (red curve in Fig.~\subfigref{fig:SD}{a}).
However, when $\Vs$ is swept in the opposite direction (blue curve), the sharp decrease of current is found at a significantly lower voltage  $\Vs\simeq 1.3$~mV.
The sharp changes in current are reproducible over several sweeps of $\Vs$, with a small variation in threshold voltages due to the stochastic nature of this process. This hysteresis cycle shows that there is a range of bias voltages, $\Vs\simeq 1.3-2.3$~mV, labeled as (II) in Fig.~\ref{fig:SD}, for which the system exhibits bistability.

We can explain the bistability regime observed using a model describing the motion of the CNT as a single vibrational mode of displacement $x(t)$, frequency $\Omega$, and effective mass $m$, whose evolution equation reads ~\cite{wachtler2019stochastic, blanter2004single, usmani_strong_2007}
\begin{equation}\label{newtoneq}
    m\ddot{x}(t) = -m\Omega^2 x(t) -  \gamma \dot x(t) - g_{\rm m} n(t){ +\xi(t)}
\end{equation}
Here $\gamma={m\Omega}/{Q}$ is the friction coefficient affecting the CNT motion and $n(t)=0,1$ is the occupation number of the dot, which is a stochastic variable,  and $\xi(t)$ is thermal Gaussian white noise with zero average and  $\langle\xi(t)\xi(t')\rangle=2\gamma k_\text{B}T\delta(t - t')$, $T$ being the CNT's temperature and $k_\text{B}$ the Boltzmann constant. Gate voltages exert a force on the CNT when the dot is occupied, i.e. $n(t)=1$. This force depends on several parameters like the distance between the dot and the gate electrodes. For small oscillations, this force can be considered constant. The constant $g_{\rm m}$ determines the strength of the coupling between the dot charge and the mechanical degree of freedom , arising from the electrostatic potential induced by the gates \cite{vigneau2021}.

The stochastic occupation number $n(t)$ undergoes Poissonian jumps  when electron tunneling occurs\cite{usmani_strong_2007, wachtler2019stochastic}. The rates of jumps from  the left  and right electrodes to the dot are given by $ \Gamma_\LR(x)f_\LR(\epsilon(x))$
and from the dot to the electrodes by
$\Gamma_\LR(x)[1- f_\LR(\epsilon(x))]$, where $\Gamma_\LR(x)$ are the tunneling rates, and $f_\LR(\epsilon(x))$ are the Fermi functions of each lead evaluated at the energy $\epsilon(x)$.The electrochemical potential of the dot, $\epsilon(x)=\epsilon_0+g_{\rm m}x$, depends on the displacement $x$ of the oscillation. The constant $\epsilon_0$ is the electrochemical potential of the dot in the absence of any mechanical motion. Assuming that the energy of the dot reaches the chemical potential $\mu_\text{S}$ at the onset of self-oscillations and transport, $\epsilon_0=eV^*_\text{S}$, where $V^*_\text{S}$ is the value of $\Vs$ at the border between regions (II) and (III) in Fig.~\ref{fig:SD}(a), i.e., $\epsilon_0=2.25$~meV. The value of $g_{\rm m}$ can be estimated as in Ref.~\cite{vigneau2021}. Notice that the tunnelling rates $\Gamma_\LR(x)$ depend on the energy of the dot and, consequently, on the displacement $x$ of the oscillations. This inhomogeneity of the tunnel barriers have been found to be a necessary condition for the occurrence of self-oscillations \cite{Laird2020, usmani_strong_2007,wachtler2019stochastic, Bennett2006Nov}.

In our experiments, the tunnelling rates are much larger than the mechanical frequency of the CNT, $\Gamma_\LR(x)\gg \Omega$. Hence, the dynamics of the random variable $n(t)$ is much faster than the motion of the CNT. Moreover, the thermal noise is small (see Supplemental Material~\cite{suppl} for a more detailed discussion of the effect of temperature in the system). This enables us to approximate the behaviour of the system through deterministic dynamics that are influenced by minor fluctuations arising from thermal noise and the random nature of the occupation number $n(t)$. The deterministic dynamics determines the stability of self-oscillations, as explained below, whereas fluctuations are responsible for their generation and decay \cite{Bachtold2020} (see also Appendices \ref{app:model} and \ref{app:decay}). If we neglect the thermal noise in Eq.~\eqref{newtoneq} and replace the random occupation number $n(t)$  by its average $\bar n(t)$ we obtain:
\begin{align}
  \ddot{x}(t) &= -\Omega^2 x(t) -  \frac{\Omega}{Q} \dot x(t) - \frac{g_{\rm m}}{m}\bar{n}(t)\label{newton2}  \\
  \dot{\bar{n}}(t) &=\Gamma_\text{in}(x(t))[1-\bar n(t)]-\Gamma_\text{out}(x(t))\bar n(t).
  \label{nave}
\end{align}
Equation \eqref{nave} is the master equation for the average occupation number with transition rates $\Gamma_\text{in}=\Gamma_\L f_\L(\epsilon)+\Gamma_\R f_\R(\epsilon)$, and $\Gamma_\text{out}=\Gamma_\L [1-f_\L(\epsilon)]+\Gamma_\R[1-f_\R(\epsilon)]$.
Eqs.~\eqref{newton2} and \eqref{nave} are deterministic and predict the appearance of self-oscillations \cite{Laird2020}. Here we analyze the stability of self-oscillations calculating the exchange of energy between the quantum dot and the CNT. To simplify the analysis, we further assume that the leads are at zero temperature, yielding  sharp Fermi functions $f_{R,L}(\epsilon)$ and that the tunneling rates are given by  $\Gamma_\LR(x)=\Gamma_\LR e^{\alpha_\LR x}$ \cite{Meerwaldt2012},  $\alpha_\LR $ being parameters that quantify the inhomogeneity of the barriers.

Self-oscillations were so far explained by proving that the last term in Eq.~\eqref{newton2} acts as negative damping that can counterbalance the energy dissipation term \cite{Laird2020}. An insightful approach to understanding the stability of self-oscillations is to consider the energy $\Delta E$ that the CNT gains in an  oscillation of a given amplitude $A$. The mechanical energy, $E(t)=m \dot x(t)^2/2+m\Omega^2 x(t)^2/2$, changes in a single oscillation of period $\tau\equiv 2\pi/\Omega$ by  
\begin{equation}
    \Delta E  = -m\int_0^\tau \left[\frac{\Omega}{Q} {\dot x(t)}^2 + \frac{g_{\rm m}}{m} { \dot x(t)}\bar n(t)\right] dt\label{eq:DeltaE},
\end{equation}
see Appendix \ref{app:DE} for details.
Notice that the device functions as an engine: the second term in Eq.\eqref{eq:DeltaE} is the energy that the  electrical charge transfers to the mechanical motion by electrostatic interaction, whereas the first term is the energy dissipated as heat through friction.

{\black To calculate $\Delta E$, it is enough to consider harmonic oscillations $x(t)=A\cos(\Omega t)$.
We solve Eq.~\eqref{newton2} for the average occupation number $\bar n(t)$ using this harmonic approximation and then insert the solution into Eq.~\eqref{eq:DeltaE} to calculate $\Delta E$ as a function of $A$ and the other relevant parameters. This approximation is accurate for a high $Q$ and a relatively small $g_{\rm m}$.}
The stability of an oscillation of amplitude $A$ is given by the sign of $\Delta E$. If the oscillation gains energy, that is, if $\Delta E >0$, then the amplitude increases. On the other hand, if $\Delta E<0$, the amplitude decreases. The magnitude of $\Delta E$ does not affect the stability of the oscillations, but it is relevant for the spontaneous generation and decay of the self-oscillations (see Appendix \ref{app:decay}).
In Fig.~\subfigref{fig:SD}{b} we show the areas where $\Delta E$ is positive (dark blue) or negative (light blue), depending on the value of the bias voltage $\Vs$ and the amplitude $A$ of the oscillation. For the rest of the parameters, we use realistic values based on a fitting of the Coulomb peaks (see Appendix \ref{app:fit}).
If the system is oscillating with an amplitude $A$ located in the dark blue region, then the oscillator gains energy in each oscillation and the amplitude increases, as indicated by the vertical arrows, until it reaches the light blue region. On the other hand, the amplitude of the oscillation decreases in the light blue region. We can also distinguish three different regions depending on $\Vs$. In region (I), $\Delta E$ is negative for any value of $A$; therefore, oscillations lose energy and fade out. In region (II), there are both positive and negative values of $\Delta E$, in correspondence with the bistability observed in Fig.~\subfigref{fig:SD}{a}. In region (III), $\Delta E$ is mostly positive, and $A$ reaches a saturation value $A\approx0.25$~nm. At the boundary between regions (II) and (III), for small values of $A$, the model also predicts unstable amplitudes. The shape of the dark blue region is given by dissipation.

Our model also provides an estimate of  $I$,
\begin{equation}
I=\frac{q}{\tau}\int_0^\tau \Gamma_\L(x(t))\Big[
f_\L(\epsilon(x(t)))-\bar n(t)\Big]dt.
 \label{eq:current}
\end{equation}
We estimate  $I$ for the $x(t)$ resulting from the saturation value of $A$. The result, plotted in Fig.~\subfigref{fig:SD}{a} (orange dashed curve), shows good agreement with the data (blue and red curves).

\begin{figure}
   \centering
    \includegraphics[scale=1]{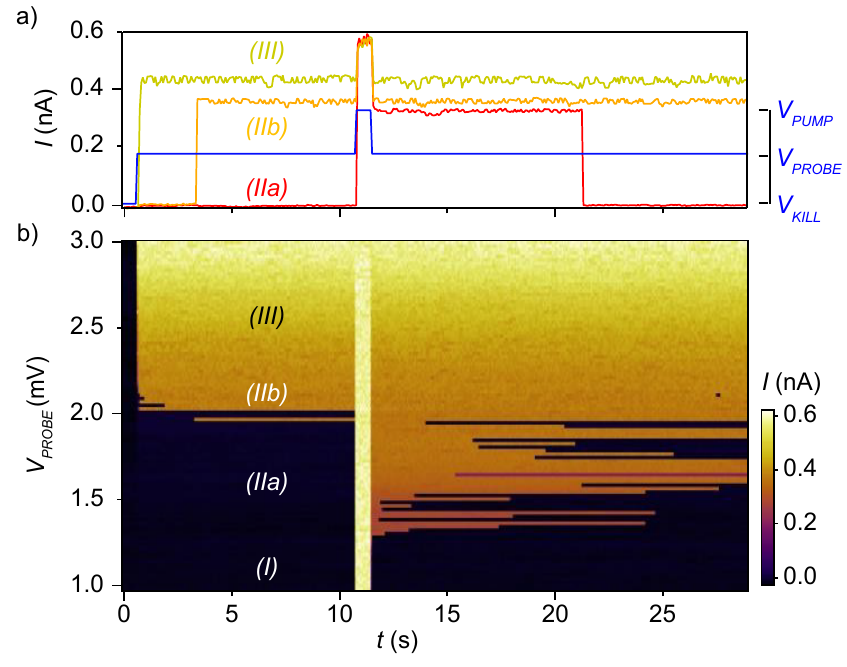}
    \caption{a) Sequence of bias voltages $\Vs$ applied (blue) and observed current during the protocol for  $\VPROBE = 1.6$~mV (IIa),  $1.98$~mV (IIb), and $2.5$~mV (III),  selected from panel b).
    The device is in the single-dot configuration (see Fig.~\subfigref{figdevice}{a}).
    $\Vs$ is initially set to $\VKILL=0$~mV to start the protocol with the CNT at rest. Then $\Vs$ is increased up to  $\VPROBE$ for a given time. self-oscillations are pumped by setting $\Vs$ to $\VPUMP=3$~mV. Finally,  the persistence of the self-oscillations is  measured by setting  $\Vs$ back to $\VPROBE$.
    b) Observed current during the protocol for different values of $\VPROBE$. We identify four regions:  (I) absence of self-oscillations; (IIa) self-oscillations observed after the pumping step and spontaneously decaying after a random time;  (IIb) self-oscillations spontaneously appearing at a bias potential $\Vs=\VPROBE$; and (III) stable self-oscillations.
    }
   \label{fig:FigStatSD}
\end{figure}

We also explore the occurrence and duration of self-oscillations in region (II). The sample is subjected to a protocol where $\Vs$ is modified in sequence as shown in Fig.~\subfigref{fig:FigStatSD}{a}.
We start the protocol with a low voltage, $\VKILL=0$~mV, for which self-oscillations are absent, and perform a rapid quench to the value $\Vs=\VPROBE$ that we want to probe. This voltage is kept constant for about 10 seconds to observe the spontaneous onset of self-oscillations revealed by a sudden increase in $I$.
$\Vs$ is then changed for about a second to a high pump bias voltage $\VPUMP=3$~mV, where self-oscillations are induced.
We then move back $\Vs$ to $\VPROBE$ for the rest of the sequence (about 20~s) to measure the persistence of the self-oscillations.
Fig.~\subfigref{fig:FigStatSD}{a} shows $I$ during the protocol for three representative regions. Sudden changes in the current indicate the presence of self-oscillations.

Fig.~\subfigref{fig:FigStatSD}{b} summarizes our experimental results for a wide range of $\VPROBE$.
We identify four different regimes associated with the three regions in Fig.~\ref{fig:SD}. For {\black $0<\VPROBE<1.3$~mV, region (I), the absence of current indicates that self-oscillations are not stable. They do not appear spontaneously at $\Vs=\VPROBE$ and vanish immediately after the pumping step.
In region (IIa), $1.3\,\, \mbox{mV}<V_\text{PROBE}<2.0$~mV}, self-oscillations do not start spontaneously but endure during a random time after being triggered by the pumping step.
In a small bias voltage range around $V_\text{PROBE} \approx 2.0$~mV, region labelled as (IIb), self-oscillations can start spontaneously after some time at $V_\text{PROBE}$.
Finally, for $V_\text{PROBE}>2~\text{mV}$, region (III), self-oscillations are always stable: they start spontaneously at $V_\text{PROBE}$ and are maintained during the whole protocol.

In the bistable regions, (IIa) and (IIb), self-oscillations appear and vanish due to fluctuations that allow the system to access areas in the stability diagram of Fig.~\subfigref{fig:SD}{b} that have the opposite sign of $\Delta E$  than that dictated by our deterministic model.  The probability of these excursions is very small but they can occur after a  large number of oscillations. For instance, a self-oscillation of frequency $\Omega/2\pi\sim 270$~MHz can last for 20 seconds or $5.4\times 10^9$ oscillations. This mechanism explains the huge separation of time scales in the device, as explained in detail in Appendix \ref{app:decay}.

\subsection{Double-dot configuration}

\begin{figure}[b]
\centering
  \includegraphics[scale=1]{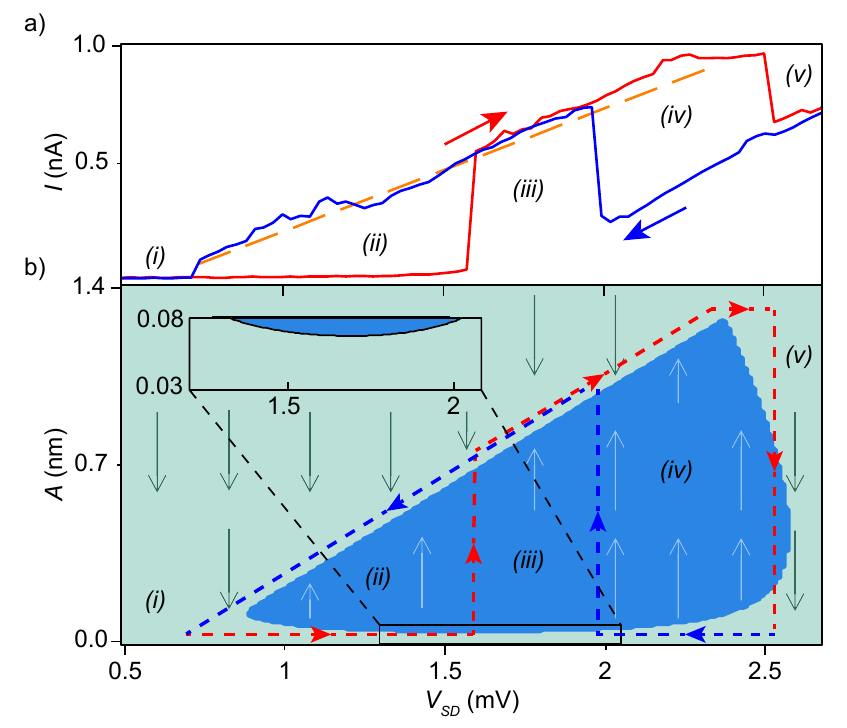}
  \centering
  \caption{
  a) Current switch hysteresis as function of $\Vs$ in the double-dot configuration, measured at the white star location in Fig.~\subfigref{figdevice}{d} ($V_\text{G3}=620$~mV, $V_\text{G5}=-120$~mV). Red and blue arrows indicate the direction of each current sweep. The dashed orange line shows the current calculated from Eq.~\eqref{eq:current}  with  $\Gamma_{\rm M} = \Omega$, $\sigma = 0.1$~nm and  amplitude  1 nm (see \ref{app:double dot}for details). The numerals (i, ii, iii, iv, v) indicate different regions in the stability diagram.
  b) Stability diagram: the dark (light) blue areas indicate $\Delta E > 0$ $(\Delta E <0)$ for different values of A and $\Vs$. Small arrows  indicate the direction in which A would change in each area given the sign of $\Delta E$. Blue and red dashed lines delimit regions (ii), (iii) and (iv). Blue and red arrows indicate a similar hysteresis cycle as that shown in panel a). Inset: zoom in on areas approaching zero amplitude. We use $\Omega/ 2\pi =270$ MHz, $g_{\rm m}=0.01$~eV$/$nm, $m =2\times 10^{-22}$ kg, and $Q = 1000$.
  \label{fig:DD}
  }
\end{figure}

We perform a similar study when the device is in the double-dot configuration by measuring the hysteresis as a function of $\Vs$ (Fig.~\subfigref{fig:DD}{a}) at the point designated by the white star in Fig.~\subfigref{figdevice}{d}. The sharp changes in current are reproducible over several sweeps of $\Vs$ for a different thermal cycle of the device (see Supplemental Material \cite{suppl}), with a small variation in threshold voltages due to the stochastic nature of this process. In the double dot configuration, the hysteretic switches define regions (i) to (iv). In region (iii) we observe current values evidencing self-oscillations, while in regions (i) and (v) these self-oscillations seem not to be present.

These current switches can be explained by our model under the assumption that the motion of the CNT mostly affects the electrochemical potential of one of the dots. This assumption is justified by the estimation of capacitive coupling of the dots to the gate electrodes which indicates that one of the dots is primarily controlled by $V_\text{G3}$, whilst the other is mainly controlled by $V_\text{G5}$. The electrochemical potential of the second dot can thus be considered aligned with $\mu_\text{D}=0$.
In this case, the system is similar to the one-dot case but replacing the transport between the dot and the rightmost electrode by tunneling between the two dots occurring when their electrochemical potentials are aligned, i.e., when $x(t)$ is close to zero; in the model, this interdot exchange is represented by the rate $\Gamma_{\rm M}$ and a width $\sigma$, $\Gamma_{\rm out}(x) \sim \Gamma_\text{M}\exp{-x^2/\sigma^2}$ (see Appendix \ref{app:double dot}).
Notice that, contrary to the case of a single dot, self-oscillations disappear for high bias voltages $\Vs$ (region (v)).
This phenomenon is difficult to explain with the model outlined above, since $\Vs$ simply determines the location of the leftmost Fermi level $\mu_\text{S}$ and should not affect the stability of self-oscillations.
However, according to Eq.~\eqref{eq:DeltaE}, self-oscillations are maintained by a correlation between the  charge  of the dot $n(t)$ and the instant velocity of the CNT,  $\dot x(t)$. This correlation could be lost due, for instance, to inelastic co-tunnelling, which is present for high values of $\Vs$, see Appendix \ref{app:double dot}.

For the double-dot case, which now considers interdot tunnelling and an inelastic contribution to the current proportional to $\Vs$ (see Appendix \ref{app:double dot}), we obtain the current shown in Fig.~\subfigref{fig:DD}{a} (orange line) and the stability diagram depicted in Fig.~\subfigref{fig:DD}{b}. The stability diagram shows good agreement with the measured current switches as a function of $\Vs$ in upwards (red) and downwards (blue) sweeps, although the boundaries of region (iii) are not precisely located due to its stochastic nature. We can also observe that the area with $\Delta E<0$ (light blue) appears close to zero $A$ in region (iii) (inset). This would indicate that the rest position of the CNT is unstable and self-oscillations can be easily induced by fluctuations, a picture which is supported by the measurements of the onset of self-oscillations in region (iii) (see Appendix \ref{app:double dot}).

\section{Conclusion}

In conclusion, we were able to construct stability diagrams that fully characterize the oscillations induced by electron tunneling in single and double quantum dot configurations. We achieve this by observing hysteresis cycles in the current flowing through the device, by using a novel protocol to probe, pump and kill these self-oscillations, and by developing a dynamical model. Our results reveal the subtleties in the coupling between mechanical motion and single electron transport, and open new venues to design autonomous motors and other types of energy transducers at the microscale.

\begin{acknowledgments}
We thank Gerard Milburn for stimulating discussions on the subject of this research. This research was supported by grant number FQXi-IAF19-01 from the Foundational Questions Institute Fund, a donor advised fund of Silicon Valley Community Foundation.
NA acknowledges the support from the Royal Society (URF-R1-191150), EPSRC Platform Grant (grant number EP/R029229/1) and from the European Research Council (ERC) under the European Union's Horizon 2020 research and innovation programme (grant agreement number 948932).
AA acknowledges the support of the Foundational Questions Institute Fund (grant number FQXi-IAF19-05), the Templeton World Charity Foundation, Inc (grant number TWCF0338) and the ANR Research Collaborative Project ``Qu-DICE" (grant number ANR-PRC-CES47).
JA acknowledges support from EPSRC (grant number EP/R045577/1) and the Royal Society.
JM acknowledges funding from the Vetenskapsr\r{a}det, Swedish VR (project number 2018-05061) and from the Knut and Alice Wallenberg foundation through the fellowship program. JT-B and JMRP acknowledge financial support from Spanish Government through Grant FLUID (PID2020-113455GB-I00)
\end{acknowledgments}

\appendix

\section{Theoretical model}\label{app:model}

In this section, we detail the theoretical model describing the coupling between the quantum dot and the mechanical oscillator. Similar models have been used in theoretical and experimental works \cite{wachtler2019stochastic, usmani_strong_2007, Laird2020}. Since the quantum dot is strongly coupled to the right and left electrodes, the system behaves classically \cite{Armour2004Mar}.
The evolution of the vertical position $x$ of the carbon nanotube (CNT) is described by an underdamped Langevin equation
\begin{equation}
    \ddot x = -\Omega^2 x - \frac{\Omega}{Q}\dot x - \frac{g_{\rm m}}{m} n + \xi.\label{Langevin}
\end{equation}
Here, $\Omega$ is the resonance frequency of the oscillator, $Q$ the quality factor, $m$  the oscillator mass, $g_{\rm m}$ the coupling constant between the oscillator and $n$ the electron occupation. The term $-\frac{g_\mathrm{m}}{m} n$ represents, precisely, the electrostatic force acting on the CNT due to the electron occupation. $\xi$ represents the environmental thermal noise, following $\langle \xi(0)\xi(t)\rangle = 2\gamma k_{\rm B}T\delta(t)$, with $T$ the environment temperature, $k_{\rm B}$ the Boltzmann constant and $\gamma = m\Omega/Q$ the damping constant. Our experiments are performed at $T = 60$ mK.

The occupation $n = 0,1$ evolves following a dichotomic stochastic process. During each time interval $dt$, the transition $n = 0\rightarrow 1$ takes place with probability $\Gamma_{\rm in}dt$, and $n = 1\rightarrow 0$ with probability $\Gamma_{\rm out}dt$. $\Gamma_{\rm in/out}$ are the energy-dependent tunnel rates defined in the main text. The average occupation $\bar n$ evolves then following the master equation
\begin{equation}
    \dot{\bar n} = \Gamma_{\rm in}[1-\bar n] - \Gamma_{\rm out}\bar n.
\end{equation}

\subsection{Mean-field approximation}

For a large quality factor $Q$ and tunnel rates $\Gamma_{\rm in/out}$, the time required for the mechanics to thermalize becomes longer than the time required by $n$ to equilibrate. In this case, both fluctuations in occupation and position are negligible, and we consider just the deterministic system of equations
\begin{align}\label{mean-field}
    \dot{\bar n} = \Gamma_{\rm in}[1-\bar n] - \Gamma_{\rm out}\bar n,\\
    \ddot x = -\Omega^2 x - \frac{\Omega}{Q}\dot x - \frac{g_\mathrm{m}}{m} \bar n,
\end{align}
which corresponds to a \textit{mean-field approximation}. In the main text, we wrote equations (2) and (3) under this approximation in order to explain the observed self-oscillations.

\subsection{Mechanical energy}\label{app:DE}

The CNT mechanical energy under the mean-field approximation is
\begin{equation}
    E = \frac{1}{2}m\dot x^2 + \frac{1}{2}m\Omega^2x^2.
\end{equation}
Taking the time derivative and using equations \eqref{mean-field} we obtain
\begin{equation}
    \frac{dE}{dt} = -m\frac{\Omega}{Q}\dot x^2 - g_\mathrm{m} \dot x\bar n.
\end{equation}
If $x(t), \bar n(t)$ is a given trajectory of Eqs.\eqref{mean-field}, the change in mechanical energy in a certain time $\tau$ is
\begin{equation}
    \Delta E = -\int_0^\tau \left[m\frac{\Omega}{Q}\dot x^2(t) + g_\mathrm{m} \dot x(t)\bar n(t) \right] dt.
\end{equation}
Equation (4) in the main text uses this expression to evaluate $\Delta E$ along a single oscillation period $\tau = 2\pi/\Omega$.

\section{Decay of self-oscillations}\label{app:decay}
The switching between states in the bistability region finds its origin in the thermal and electric noise, see Eq.~\eqref{Langevin}. In order to clarify this, we expose the bistability in terms of the double-well problem \cite{gardiner, usmani_strong_2007, blanter2004single} in the single-dot configuration.
We now consider a long timescale $T_A\gg 1/ \Omega$ where the amplitude of oscillation will vary with time. In this timescale, we consider an effective Langevin equation for mechanical energy,
\begin{equation}
   \frac{dE}{dt} = f(E) + \xi_E,\label{dEdt}
\end{equation}
where $f(E)=\Delta E/\tau$ is the average increment of energy per unit of time, and $\Delta E$ is given by Eq.~(4) of the main text. Note that the value of $\Delta E$ depends on the amplitude of the oscillations, and therefore on the mechanical energy, $E = m\Omega^2 A^2/2$. $\tau = 2\pi/\Omega$ is the period of the oscillation. $\xi_E$ is a noise containing both electrical and thermal fluctuations, and will depend on the actual value of the energy.
Equation \eqref{dEdt} can be written as a Kramers equation considering an effective ``potential'' given by
\begin{equation}
    U(E) = -\frac{1}{\tau}\int_0^E \Delta E(E') dE',
\end{equation}
then,
\begin{equation}
   \frac{dE}{dt} = -\frac{dU}{dE} + \xi_E.
\end{equation}

\begin{figure}[ht]
    \centering
    \includegraphics{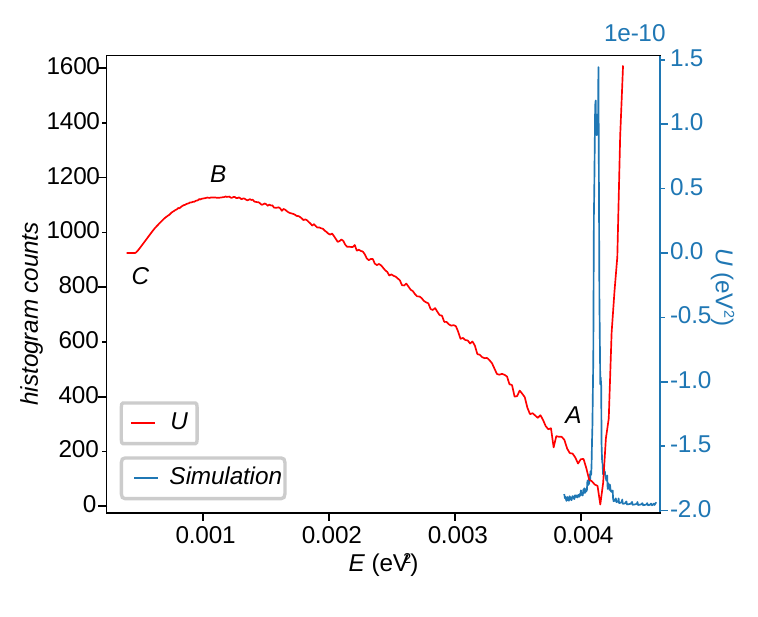}
    \caption{Numerical computation of the effective potential $U(E)$ (red) and stochastic simulation of a single trajectory (blue) as a function of the mechanical energy for the one dot-configuration. The sharp peak in $U$ at the point A is due to the end of the borders of the conduction region of the device. The letters A, B, C indicate the stable oscillation points (A, C) and the tipping point between both regimes (B). The parameters are the same as in the main text, $\Omega/2\pi=270$ MHz, $g_{\rm m}=0.01$~eV$/$nm, $m =2\times 10^{-22}$ kg, $Q = 1000$, $\Gamma_\L=400~\text{ GHz}$, $\Gamma_\R=18~\text{ GHz}$,  $\alpha_\L=4~\text{ nm}^{-1}$, and  $\alpha_\R=0$.}
    \label{fig:potential}
\end{figure}
In Fig.~\ref{fig:potential}, we represent the effective potential $U(E)$ (red line), evaluated using harmonic trajectories $x(t) = A \cos \Omega t$. Here, the problem is explicitly analogous to the typical double-well potential in energy space. In that figure, self-oscillations correspond to the right well, A, with a certain energy associated. The other well, C, corresponds with the motionless state. Switching between both states will happen when a fluctuation puts the system over the barrier B.

We use equation \eqref{Langevin} to simulate one trajectory in the self-oscillation state for $10^4$ oscillation periods. The histogram of this trajectory is included in Fig.~\ref{fig:potential} (blue). This histogram is sharply peaked in the point A, and therefore a very large number of mechanical periods are required for the self-oscillations to vanish.

According to this, the self-oscillation duration is given by the shape of the barrier. In the region (II) of figure 2 in the main text, the size of the barrier increases with the bias voltage $\Vs$, generating longer and longer self-oscillations in that region. This increase in duration is observed in figure 2 of the main text along region (IIa).

\section{Fitting of the Coulomb peak and tunneling rates}\label{app:fit}

\begin{figure}[ht]
  \includegraphics{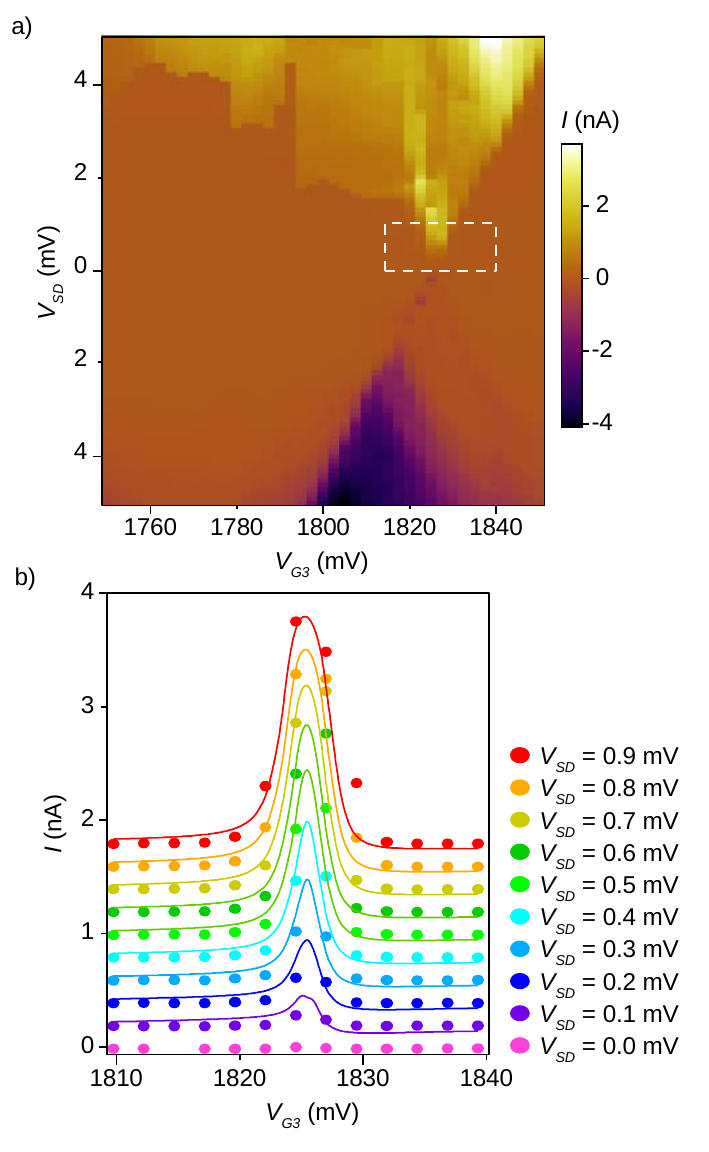}
  \caption{Coulomb diamond and fit.
  a) Zoom on the Coulomb diamond from Fig.~1d of the main text. The white dashed rectangle represent the area from which are obtained the cut of panel b).
  b) Set of cuts of panel a) at different bias (circles) and fitting curves (lines). For each bias, the cuts and fit are vertically offset by 0.2~nA from the previous  bias. The fitting parameters of the all set are: $\GL$ = 400~GHz, $\GR$= 15~GHz, $\alpha_\L/g=0~\text{eV}^{-1}$, $\alpha_\R/g=400~\text{eV}^{-1}$ and $V_0=1.826$~V. The lever arm 0.18~eV/V is obtained from panel a).
  \label{fig:SuppCoulombPeakFit}
of  }
\end{figure}

We fit the Coulomb diamond in the single dot configuration to estimate the tunnelling rates parameters $\G_\LR$ and $\alpha_\LR$. The fitting procedure is similar to our previous work~\cite{vigneau2021} with the addition of energy-dependent tunnelling rates, represented by the parameter $\alpha_\LR$, inspired from Ref.~\onlinecite{Meerwaldt2012}. The rates of charges tunnelling in and out between the left (L) or right (R) lead and the quantum dot are given by the expression:
\begin{subequations}\label{eq:tunnel rates}
\begin{align}
  &\G_\LR^\In(\epsilon)=\G_\LR\exp(\alpha_\LR\epsilon/g_{\rm m}) ~\rho_\LR(\epsilon),\\
  &\G_\LR^\Out(\epsilon)=\G_\LR\exp(\alpha_\LR\epsilon/g_{\rm m}) \left  (1-\rho_\LR(\epsilon)\right),
\end{align}
\end{subequations}
where, as in the main text, $\epsilon$ is the energy of the dot and $f_\LR$ are the Fermi distribution.

The overlap between the density of states of the quantum dot and left/right reservoirs is given by:
\begin{equation}
\rho_\LR(\epsilon)=\frac{1}{2}+\frac{1}{\pi}\arctan \left( \frac{2(\upmu_\LR-\epsilon)}{\hbar\Gb} \right ).
\end{equation}
Finally, the current across the quantum dot is:
\begin{equation}
I(\epsilon) = e  \frac{\GL^\In(\epsilon)  \GR^\Out(\epsilon)-\GR^\In(\epsilon)  \GL^\Out(\epsilon)}{\Gb}.
\label{equ:current}
\end{equation}

We fit the experimental the Coulomb diamonds in Fig.~\subfigref{fig:SuppCoulombPeakFit}{a} by cutting it into multiple Coulomb peak with bias ranging from $\Vs=0$~mV to $\Vs=0.9$~mV (Fig.~\subfigref{fig:SuppCoulombPeakFit}{b}). We fit these Coulomb peaks with a unique set of parameters to obtain $\GL$ = 400~GHz, $\GR$= 15~GHz, $\alpha_\R/g_{\rm m}=0~\text{eV}^{-1}$ and $\alpha_\L/g_{\rm m}=400~\text{eV}^{-1}$.

To take into consideration the voltage drop at the IV converter internal resistance (100~k$\Omega$), we reduce the bias of the fit by $\Vs^\text{Corr}(V_\mathrm{G1})=\Vs-I(V_\mathrm{G3})R_\mathrm{s}$. Because the density of the point of the measurement is not sufficient, we calculate $\Vs^\text{Corr}$ from a first fit of the Coulomb peaks. We then fit again the peaks considering the corrected bias voltage. The contribution of this correction on the final result is small.

\section{Double-dot case}\label{app:double dot}

As exposed in the main text, the gate voltages inducing the configurations are $V_\text{G3}$ and $V_\text{G5}$, and then one of the dots is located close to the center of the CNT and can moves as in the single-dot configuration, whereas the second one is close to the rightmost end of the CNT, remaining static. This system is similar to the one-dot case but replacing the transport between the moving dot and the right lead by an effective tunnelling rate with the form  $\Gamma_{\rm dot-dot}(x) = \Gamma_\text{M}\exp{-x^2/\sigma^2}$. The width $\sigma$ represents the effects of the thermal noise on the moving dot.

However, notice that, contrary to the case of a single dot, self-oscillations disappear for high bias voltages $\Vs$, see main text. This happens as a result of the inelastic current, promoting transport even between misaligned dots \cite{hanson2007spins,van2002electron}.
In (iii, iv, v) we will calculate inelastic transport in a completely effective way, including homogeneous transport rates $\Gamma^{\rm ie}(\Vs)$ in the master equation,
\begin{align}
\Gamma_{\rm in}(x)&=\GL(x)f_\L(\epsilon(x)) + \Gamma^{\rm ie}(\Vs) \\
\Gamma_{\rm out}(x)&=\GL(x)[1-f_\L(\epsilon(x))]+\Gamma_{\rm dot-dot}(x)+ \Gamma^{\rm ie}(\Vs),
\end{align}
and obtaining their value from the motionless state. In regions (i, ii) we state $\Gamma^{\rm ie}(\Vs) = 0$ for every $\Vs$.

\begin{figure*}[htb!]
    \centering
    \includegraphics{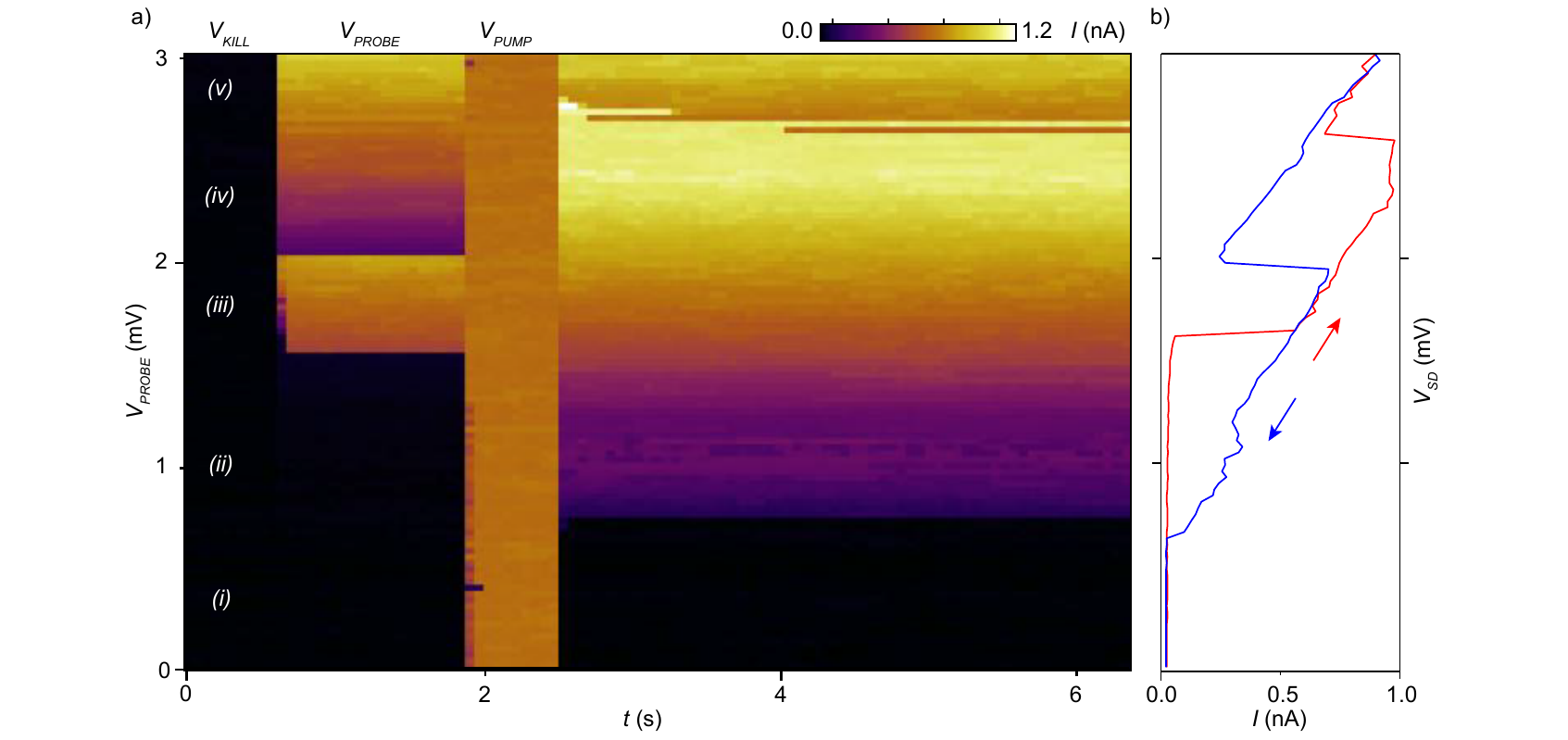}
    \caption{
  a) Current measured in the double-dot configuration by repeating the sequence shown Fig.\ref{fig:FigStatSD} for different $\VPROBE$. First, any self-oscillations are killed setting $\Vs$ to $\VKILL=0$~mV, then $\Vs$ is set to the target voltage $\VPROBE$. About 1.5~s later, $\Vs$ is set to $\VPUMP=1.8$~mV to start self-oscillations for about 0.5~s. Finally, the voltage is set back to the target $\Vs$. The different behaviours are classified in five areas: (i) absence of self-oscillations, (ii) self-oscillations observed after the pumping step, (iii) self-oscillations spontaneously appearing, (iv) self-oscillations observed after the pumping step and spontaneously decaying after a random time, (v) absence of self-oscillations.
  b) IV measurement with the two sweep direction in the configuration of panel b) showing the current switch hysteresis; same as Fig.~4(a) of the main text.
    \label{fig:DD_stats}}
\end{figure*}

For the double-dot configuration, we have implemented a similar protocol as for the single-dot configuration in Fig. 3 of the main text (Fig.~\subfigref{fig:DD_stats}{a}). The measurement is performed at the coordinates of the white star in Fig. 1e of the main text, in the same configuration as in Fig. 4 of the main text. The pump voltage $\VPUMP=1.8$~mV is chosen in the area where the self-oscillations are always on (Fig.~\subfigref{fig:DD_stats}{b}).

From the current measurement, depicted in Fig.~\subfigref{fig:DD_stats}{b}, we identify different behaviours depending on $\Vs$. In (i) and (ii), self-oscillations never start spontaneously, whereas in (iii) they start spontaneously just after the kill step.  After the pump step, in regions (i) and (v) self-oscillations stop immediately. In regions (ii) and (iv) the triggered self-oscillations are sustained for a duration too long to be observed in the time frame of this experiment. In the border between regions (iv) and (v) the self-oscillation time decays in a similar way as observed in the single-dot configuration.

\newpage
\onecolumngrid
\setcounter{figure}{0}
\setcounter{equation}{0}
\renewcommand{\thefigure}{S\arabic{figure}}

\end{document}


\title{Supplemental Material: Stability of long-sustained oscillations induced by electron tunneling}

\author{Jorge Tabanera-Bravo}
\affiliation{Dept.~Estructura de la Materia, F\'isica T\'ermica y Electr\'onica and GISC, Universidad Complutense de Madrid. 28040 Madrid, Spain}

\author{Florian Vigneau}
\affiliation{Department of Materials, University of Oxford, Oxford OX1 3PH, United Kingdom}

\author{Juliette Monsel}
\affiliation{Department of Microtechnology and Nanoscience (MC2), Chalmers University of Technology, S-412 96 G\"oteborg, Sweden\looseness=-1}

\author{Kushagra Aggarwal}
\affiliation{Department of Materials, University of Oxford, Oxford OX1 3PH, United Kingdom}

\author{L\'ea Bresque}
\affiliation{Univ.~Grenoble Alpes, CNRS, Grenoble INP, Institut N\'eel, 38000 Grenoble, France}

\author{Federico Fedele}
\affiliation{Department of Materials, University of Oxford, Oxford OX1 3PH, United Kingdom}

\author{Federico Cerisola}
\affiliation{Department of Materials, University of Oxford, Oxford OX1 3PH, United Kingdom}
\affiliation{Physics and Astronomy, University of Exeter, Exeter EX4 4QL, United Kingdom}

\author{G.A.D. Briggs}
\affiliation{Department of Materials, University of Oxford, Oxford OX1 3PH, United Kingdom}

\author{Janet Anders}
\affiliation{Physics and Astronomy, University of Exeter, Exeter EX4 4QL, United Kingdom}
\affiliation{Institut f\"ur Physik und Astronomie, Potsdam University, 14476 Potsdam, Germany}

\author{Alexia Auff\`eves}
\affiliation{Univ.~Grenoble Alpes, CNRS, Grenoble INP, Institut N\'eel, 38000 Grenoble, France}

\author{Juan M.R. Parrondo}
\affiliation{Dept.~Estructura de la Materia, F\'isica T\'ermica y Electr\'onica and GISC, Universidad Complutense de Madrid. 28040 Madrid, Spain}

\author{Natalia Ares}
\email{natalia.ares@eng.ox.ac.uk}
\affiliation{Department of Engineering Science, University of Oxford, Oxford OX1 3PJ, United Kingdom}

\maketitle

We show here additional measurements complementing our results from the main text.

\section{Stability diagram}

\begin{figure*}[h!]
  \includegraphics{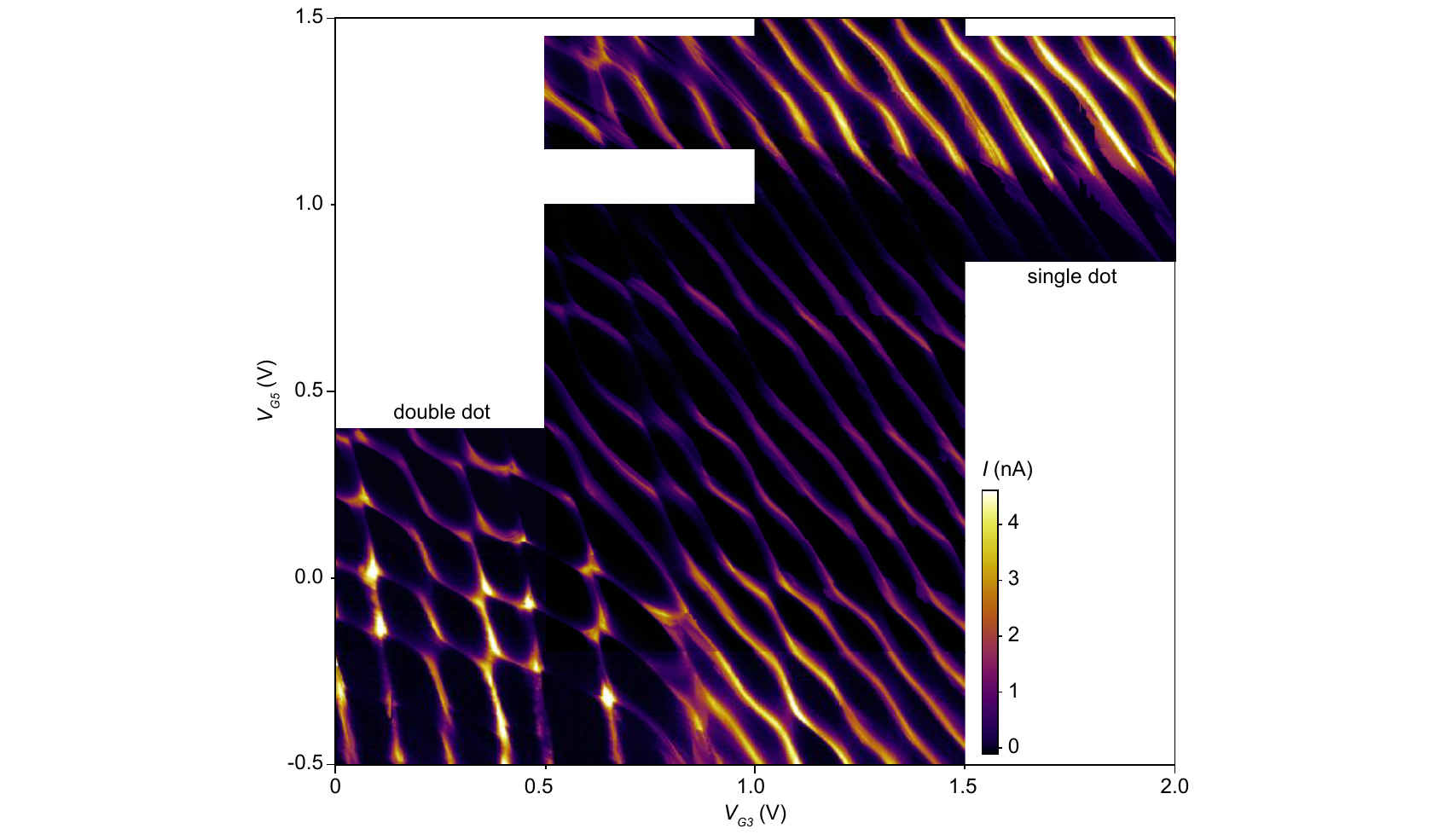}
  \caption{Stability diagram containing the single quantum dot  and double quantum dot regime measured at $\Vs=3$~mV.
  \label{fig:SuppBigStabilityDiagram}
  }
\end{figure*}

In Fig. \ref{fig:SuppBigStabilityDiagram} we show a large measurement of the stability diagram of the carbon nanotube going from the single dot to the double dot configurations during the cool down run of the main paper results.

\section{Self Oscillations in double quantum dots}

Now, we present additional results obtained with the same device but during a different cool down. The device have been endured a thermocycle at 300~K between the results shown in the main text and the one presented in the following.

\begin{figure*}[ht]
  \includegraphics{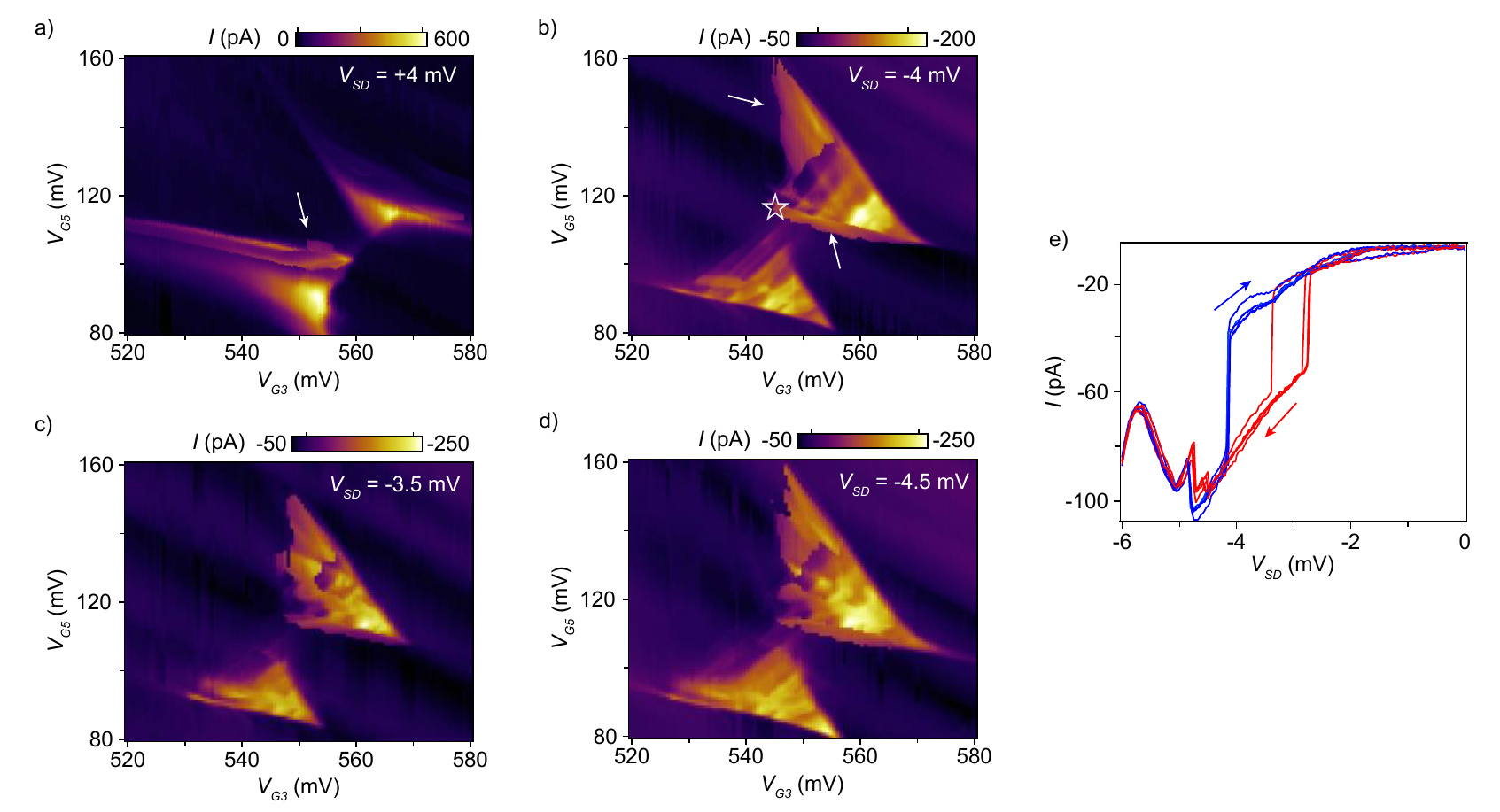}
  \caption{Results obtained for the same device but during a different cool down than the results presented in the main text. 
  a,b) Bias triangles measured on the double quantum dot regime for $\Vs=+4$~mV (a) and $\Vs=-4$~mV. Self-oscillations are visible on the edge of one of the triangles highlighted by the arrow. (c,d) Measurements of self-oscillations in the same conditions as panel (b)but with $\Vs=-3.5$~mV and $\Vs=-4.5$~mV for panel (c) and (d) respectively. e) Self-oscillation hysteresis measured at the position of the white star in panel (b). We report five sweeps forward sweep (blue), and five backward sweep (red). The sweep direction is indicated by arrows.
  \label{fig:SuppDQDDec}
  }
\end{figure*}

In Fig.~\subfigref{fig:SuppDQDDec}{a} and Fig.~\subfigref{fig:SuppDQDDec}{b} we show the measurement of a bias triangle with positive and negative bias voltages ($\Vs = \pm4$~mV) showing that the device is in the double quantum dot configuration.  We measure the hysteresis of the self-oscillations (Fig.~\subfigref{fig:SuppDQDDec}{e}) at the coordinates of the white star in Fig.~\subfigref{fig:SuppDQDDec}{b}, located in the edge of the self-oscillation area. The hysteresis is measured multiple time in a row and show a bistability region from $\Vs = -4$~mV to $\Vs = -2$~mV. 
Following the homogeneous area from the white star, evidence of self-oscillations are visible one particular edge of one of the triangle in both bias polarity. This edge correspond to the chemical potential of one of the two dots (the same in both polarity) align with the chemical potential of the contact lead of its side~\cite{hanson2007spins}. Self-oscillations are also observed on the same side of the bias triangles for bias voltages ranging from -3.5~mV to -4.5~mV, see Fig.~\subfigref{fig:SuppDQDDec}{b,c,d}.

\begin{figure*}[ht]
  \includegraphics{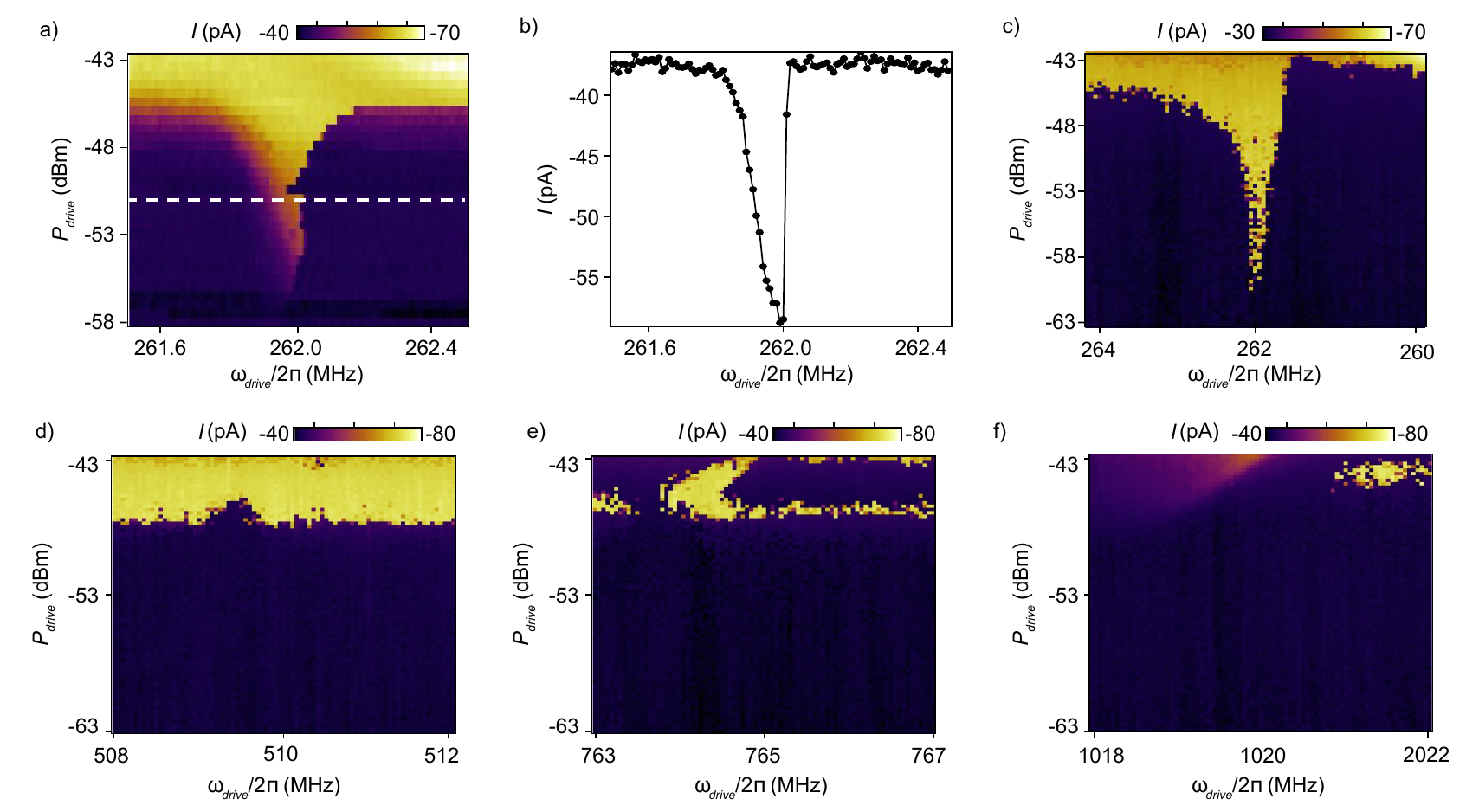}
  \caption{ Mechanical quality factor and mechanical assisted self-oscillations:
  a) Current measured as a function of the rf excitation at frequency $\omega_\mathrm{e}/2\pi$ and power $P_\mathrm{drive}$. The source-drain bias is set to $\Vs= -4$~mV in a region without self-oscillations, (see Fig.~\subfigref{fig:SuppDQDDec}{e}). The white dashed lines indicate the position of the line cut in panel (b).
  b) Line-cut of panel (a) at $P_\mathrm{drive}=-51$~dBm (white dashed line). From the width of the dip at half minimum, we estimate a mechanical quality factor of $Q\approx$ 2000.
  c) Current measured as a function of the rf excitation at frequency $\omega_\mathrm{e}/2\pi$ and power $P_\mathrm{drive}$. Here $\Vs= -3$~mV is set in the middle of the bi-stable region (see Fig.~\subfigref{fig:SuppDQDDec}{e}) marking the presence of self-oscillations. Mechanically assisted self-oscillations are induced by applying a continuous rf-excitation $\omega_\mathrm{e}/2\pi$ near the mechanical resonance frequency. The measurement is performed at a location marked by the star marker in Fig.~\subfigref{fig:SuppDQDDec}{b}. For each pixel self-oscillations are initially suppressed by setting $\Vs= 0$~mV for $0.1$~s, and then set to $-3$~mV while measuring the current for $0.1$~s. The bright pixels indicate high negative current and mark the points where self-oscillations are present. 
  d-f) Same measurement as in a) but at frequencies $\omega_\mathrm{e}/2\pi$ attributed to the second, third and fourth mechanical modes. No mechanically assisted self-oscillations are observed for these modes.
  \label{fig:Supp_Mechanics}
  }
\end{figure*}

To evidence the mechanical nature of the current switch that we associate with self-oscillations, we use a radio-frequency (rf) excitation, applied to one of the gate electrodes, to drive the carbon nanotube and excite self-oscillations. This is shown in Fig.~\ref{fig:Supp_Mechanics}.
First, we reveal the mechanical resonance frequency $\Omega=262$~MHz in Fig.~\subfigref{fig:Supp_Mechanics}{a} by measuring the current as a function of the rf drive frequency $\omega_\mathrm{drive}/2\pi$ and power $P_\mathrm{drive}$. The mechanical quality factor, $Q \approx$  2000, is estimated from the width at a half minimum of the peak in current displayed in Fig.~\subfigref{fig:Supp_Mechanics}{b} 

We then apply the rf excitation while the device is in a bistability region (white arrow in Fig.~\subfigref{fig:SuppDQDDec}{b}, $\Vs =-3$~mV). For each pixel in  Fig.~\subfigref{fig:Supp_Mechanics}{c}, the bias voltage $\Vs$ is initially set to $\Vs =0$~mV for 0.1~s to kill any self-oscillations and then set back to $\Vs =-3$~mV for an other 0.1~s, when the current is measured. Most of the times, the self-oscillations consistently restart when the drive frequency is near $\Omega$.
This result evidences a mechanical pumping of the self-oscillation at much reduced power $P_\mathrm{drive}$ when the frequency of the excitation $\omega_\mathrm{drive}$ is close to $\Omega$. 

\section{Statistics for the duration of self-oscillations}

\begin{figure*}[ht]
  \includegraphics[scale = .5]{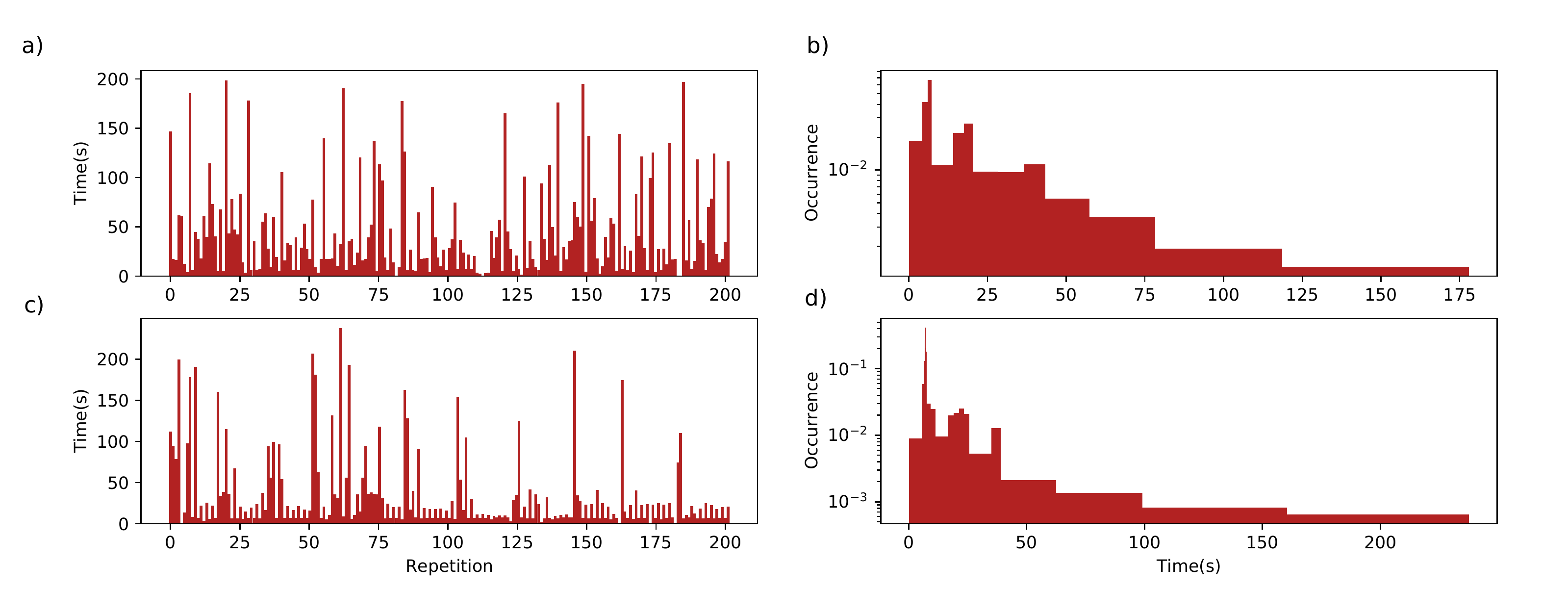}
  \caption{
  a,b) 200 measurements in a row of the duration of self-oscillations for two different values of $\Vs$. 
  For each vertical line, the self-oscillations are first excited by setting the bias voltage to $\Vs=-4$~mV for about 5 s then the bias is set to the target voltage $\Vs= -3$~mV (a) or $\Vs= -3.2$~mV (c). Then we measure the current at a rate of about one point per second. We detect the stop of the self-oscillation when the current falls below a pre-determined threshold. The top of each row indicates the duration of the self-oscillations.
  b,d) Histograms of the duration of self-oscillation for (a) and (c), respectively.
  \label{fig:Supp_Statistics}
  }
\end{figure*}

In Fig.~\ref{fig:Supp_Statistics}, we show a statistical measurement of the self-oscillation time in the same experimental run as Fig.~\ref{fig:SuppDQDDec} and \ref{fig:Supp_Mechanics}. First, we prepare the device in the bistability region at the gate voltage coordinates corresponding to the white star in Fig.~\subfigref{fig:SuppDQDDec}{b}. Then, we excite self-oscillations for about 5~s by setting the bias voltage to $\Vs =-4$~mV, setting for which self-oscillations are always ON (see Fig.~\subfigref{fig:SuppDQDDec}{c}). We then change the bias voltage to probe the bistability region ($\Vs =-3$~mV for Fig.~\subfigref{fig:Supp_Statistics}{a} and -3.2~mV for Fig.~\subfigref{fig:Supp_Statistics}{c}) and measure the current with a rate of approximately 1 point per second. When the current falls below a certain threshold, the measurement stops and the sequence restarts to the next column (saving significant measurement time). We ran 200 sequences for each figure (\subfigref{fig:Supp_Statistics}{a} and \subfigref{fig:Supp_Statistics}{c}), displaying a series of vertical lines from which we can extract the duration of the self-oscillations.

We plot a histogram of the duration of the self-oscillations in Fig.~\subfigref{fig:Supp_Statistics}{b} and \subfigref{fig:Supp_Statistics}{d}.
We note a pattern, particularly visible in Fig.~\subfigref{fig:Supp_Statistics}{c}, where long lived self-oscillations are followed by a rather repetitive number of short lived self-oscillations. In time domain, this correspond to an approximate 8 min period, in which the first 3 min shows a few long lived self-oscillations and the next 5 min a series of short lived self-oscillations. This regular variability of self-oscillations suggests an influence of external sources which could not be identified.

\section{Temperature dependence}

\begin{figure*}[ht]
  \includegraphics{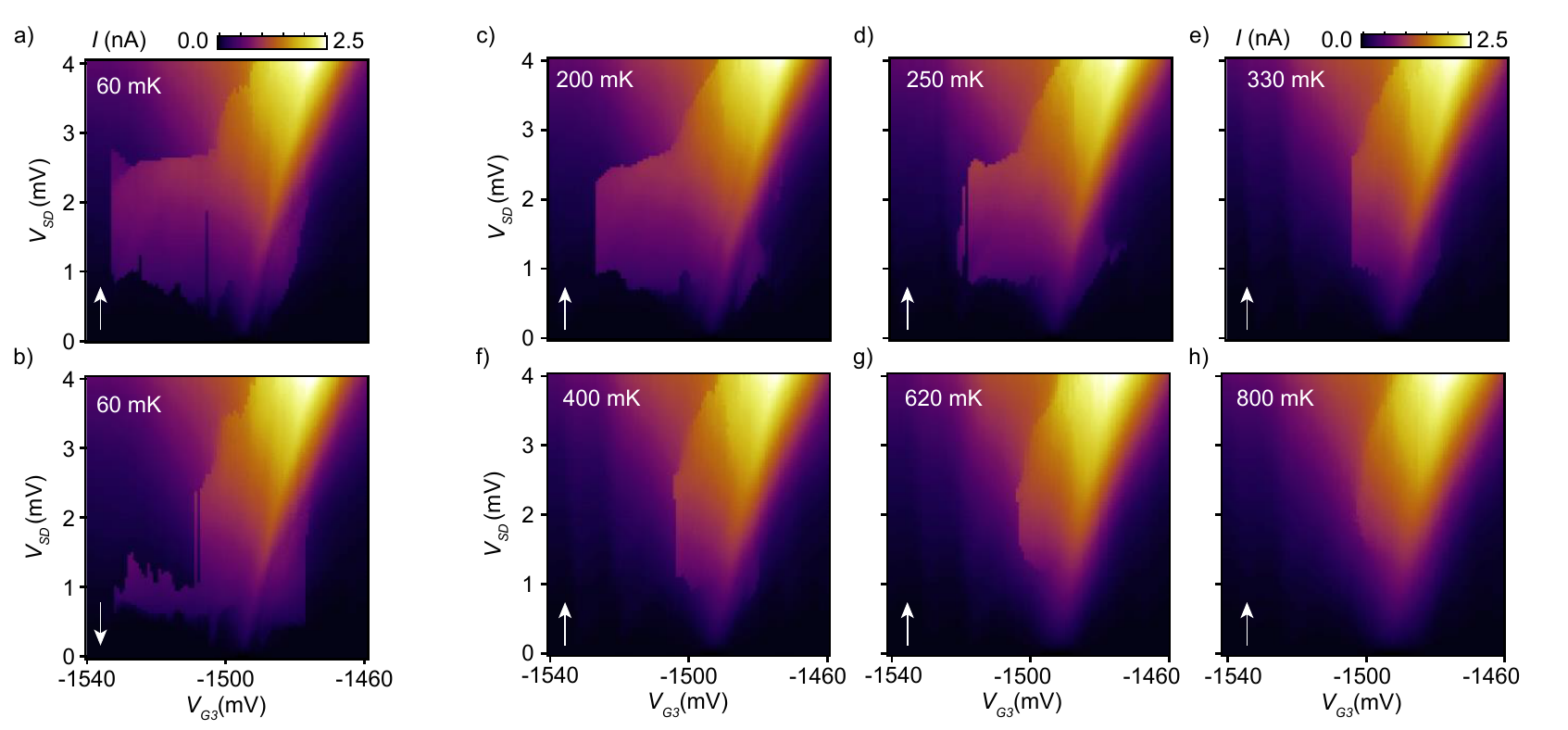}
  \caption{ Measurements of self-oscillations in a Coulomb diamond for different temperatures. The leftmost panel was measured at base temperature, 60~mK, with two different voltage bias sweep directions indicated by white arrows. They reveal an area of self-oscillations and another of bistability showing self-oscillations in one of the sweep directions. 
  In the other panels, the Coulomb diamond is always measured with the same bias sweep direction but with increasing temperatures, up to 820~mK. The self-oscillation area progressively vanishes as the temperature increases.
  \label{fig:Supp_Temperature}
  }
\end{figure*}

In yet another cool down run, we measure self-oscillations in the Coulomb blockade regime under various temperature conditions. 
In Fig.~\ref{fig:Supp_Temperature}, we show the positive side of a Coulomb diamond, showing a region where self-oscillations are  evident. The self-oscillations are first measured at 60~mK (base temperature of the cryostat) under both bias sweep directions (two leftmost panels in Fig~\ref{fig:Supp_Temperature}) in order to highlight a bistability region. Then the same Coulomb diamond was measured at increasing cryostat temperatures up to a maximum of 820~mK (other panels in Fig.~\ref{fig:Supp_Temperature}). Note that a 20~mK uncertainty in the device temperature should be considered to take into account the thermalization of the chip. We saw that the self-oscillation area decreases progressively, starting from the regions further away from the Coulomb diamond, as the temperature rises.

The damping of self-oscillations at high temperatures can be explained by the dispersion of the dynamics due to fluctuations. This magnitude can be estimated as $\sigma_x^\mathrm{therm} \sim \sqrt{k_{\rm B}T/m\Omega^2}$, where $k_{\rm B}$ is the Boltzmann constant and $T$ is the cryostat temperature. For $T = 60$~mK, it takes the value $\sigma_x^\mathrm{therm}\simeq 3.7 \times 10^{-2}$ nm; since this value is orders of magnitude smaller than the amplitude of the self-oscillations, its effect on the oscillations is negligible. However, for $T = 1$ K the dispersion is $\sigma_x^\mathrm{therm}\simeq 0.14$ nm. In this regime, the dispersion in the dynamics due to temperature is of the size of the self-oscillations, preventing their appearance.